\documentclass[conference]{IEEEtran}

\usepackage{myStyleIEEE}

\newtheorem{definition}{Definition}
\newtheorem{theorem}{Theorem}

\usepackage[margin=0.65in]{geometry}
\usepackage{booktabs}
\usepackage{tabularx}
\usepackage{amsmath}

\IEEEoverridecommandlockouts

\begin{document}

\title{Grid-Forming Characterization in DC Microgrids}

\renewcommand{\theenumi}{\alph{enumi}}

\newcommand{\jovan}[1]{\textcolor{magenta}{$\xrightarrow[]{\text{J}}$ #1}}
\newcommand{\ognjen}[1]{\textcolor{pPurple}{$\xrightarrow[]{\text{O}}$ #1}}

\author{\IEEEauthorblockN{Jovan~Krajacic\IEEEauthorrefmark{1}\IEEEauthorrefmark{2},
        Ognjen~Stanojev\IEEEauthorrefmark{1},
        Mario~Schweizer\IEEEauthorrefmark{1},
        Orcun~Karaca\IEEEauthorrefmark{1},
        Gabriela~Hug\IEEEauthorrefmark{2},
        Vladan Lazarević\IEEEauthorrefmark{1}}
        \IEEEauthorblockA{\IEEEauthorrefmark{1} ABB Corporate Research Center, Switzerland}
        \IEEEauthorblockA{\IEEEauthorrefmark{2} EEH - Power Systems Laboratory, ETH Zurich, Switzerland}
        \{jkrajacic, hug\}@eeh.ee.ethz.ch, \{ognjen.stanojev, mario.schweizer, orcun.karaca, vladan.lazarevic\}@ch.abb.com
\thanks{\textit{Corresponding Author:} Ognjen Stanojev, ognjen.stanojev@ch.abb.com}
}

\maketitle
\IEEEpeerreviewmaketitle

\begin{abstract}
DC microgrids are converter-based electrical networks that are increasingly being used in various applications, including data centers and industrial distribution systems. A central challenge in their operation is maintaining the DC-bus voltage within predefined limits while ensuring overall system stability. Although a wide variety of converter control algorithms has been proposed to achieve these objectives, the literature lacks a clear and physically interpretable framework for evaluating their effectiveness and for classifying and comparing them. Moreover, the grid-forming versus grid-following distinction that exists in AC systems has largely been unexplored in DC microgrids. To address this gap, this paper introduces three novel impedance-based indices that can be used to quantify the voltage-forming and current-forming behavior of a converter. The indices also provide a basis for defining the desired converter behavior that yields superior DC-bus voltage regulation performance. Simulation results illustrate the application of the framework to several representative control strategies and highlight the strengths and limitations of these control algorithms.
\end{abstract}


\section{Introduction} \label{sec:intro}
The adoption of DC microgrids has accelerated across various sectors, including commercial buildings, data centers, and industrial distribution systems, due to their cost-effectiveness, transmission efficiency, and compatibility with renewable sources. Despite their advantages, the converter-based nature of DC networks introduces control and protection challenges~\cite{Dragicevic2016}. The central operational problem is to regulate the DC voltage within prescribed limits while guaranteeing stability in the presence of tightly controlled power-electronic converters and fast load transients. Furthermore, the need for proportional load sharing among parallel source converters is critical to prevent individual unit overloading and to ensure power balance~\cite{Guerrero2021}. As a result, the development of decentralized control strategies has become a primary focus to ensure autonomous and reliable operation under varying loads.

Within the decentralized control paradigm, droop‑based schemes are widely adopted in both research and standards~\cite{dcindustrie2_2024}. Their operation is grounded in proportional control, resulting in simple and modular implementations. Two principal categories of droop control exist~\cite{Gao2017, CompareDroop2017,Wang2018}: current‑mode (I–V) and voltage‑mode (V–I). The former computes the current reference from the measured output voltage, whereas the latter derives the voltage reference from the measured output current. Several enhanced variants of droop control have been proposed to introduce additional control features or to improve the performance of the two basic droop control schemes. Enhancements to V--I droop control commonly employ virtual capacitance and damping‑current injection to emulate inertia and damping~\cite{Zhu2020, 10621669}. Similar improvements have also been considered for I--V droop control~\cite{8096714}. Another means of providing inertia and damping in DC microgrids is through a distinct category of converter control known as the Virtual DC Machine (VDCM), which aims to emulate the electromechanical dynamics of a DC machine. Implementations of VDCM control include power‑based formulations~\cite{Tan2016VDCM}, torque‑based designs~\cite{Lin2021VirtualInertia}, and variants without power or torque signals~\cite{Na2022PowerLoopFree}.

Assessing and comparing the performance of the different converter control strategies is typically carried out by evaluating the converter’s transient voltage response, which requires a detailed simulation setup and implementation of each control scheme. In contrast, the output impedance of the converter can be obtained analytically and is widely used as a foundation for stability analysis in DC microgrids \cite{lazarevic_dc_microgrids, Gao2017, he2025small}. However, it has been used only to a limited extent for comparing and evaluating control strategies. Among the few examples, \cite{CompareDroop2017} examined the output impedance of I--V and V--I droop by analyzing the influence of controller parameters on the resulting impedance transfer functions, indicating that I--V droop offers a better stability margin compared to V--I droop control, especially under small virtual resistance values. Alternatively, \cite{liu2020} established a correlation between output impedance magnitude overshoots and transient voltage excursions. Despite these insights, the analysis of the output impedance remains underutilized and not fully defined as a general performance metric for source converter behavior, leaving the comparative assessment of droop algorithms without a unified framework.

Another significant gap exists in the research on DC microgrids. While numerous control algorithms have been proposed, as discussed above, a fundamental question remains unresolved: what constitutes the desired converter control behavior, or more precisely, what is the ideal converter response to disturbances in the grid? In AC microgrids, the desired control behavior of source converters is commonly associated with ideal voltage source characteristics and is generally referred to as \textit{grid-forming} control~\cite{Alican2025}. Furthermore, several recent works have proposed ways to quantify the grid-forming capabilities~\cite{Debry2019,zhuang2025}. In contrast, for DC microgrids, there is neither a universally accepted definition nor a quantitative characterization of (desired) grid-forming behavior.

To address the identified research gaps, this paper makes the following contributions. Firstly, we introduce three output‐impedance-based indices that quantitatively characterize the voltage-forming behavior of a converter, defined as the sensitivity of the output voltage to variations in the output current, and its current-forming behavior, defined as the sensitivity of the converter current to variations in the output current. In contrast to existing grid-forming and grid-following classifications developed for AC systems \cite{Debry2019,zhuang2025}, the proposed indices distinguish among voltage forming, current forming or current following, and disturbance amplifying behavior, depending on their numerical values. This provides a unified and physically interpretable framework for comparing converter control algorithms without requiring detailed knowledge of the surrounding DC microgrid. Secondly, the indices establish a basis for defining converter behavior that achieves desired DC bus voltage regulation, thereby enabling a clear and quantitative definition of grid-forming behavior in DC microgrids.

\section{Control of Source Converters in DC Microgrids} \label{sec:control_algorithms}
\begin{figure}[b!]
    \centering
    \includegraphics[width=0.480\textwidth]{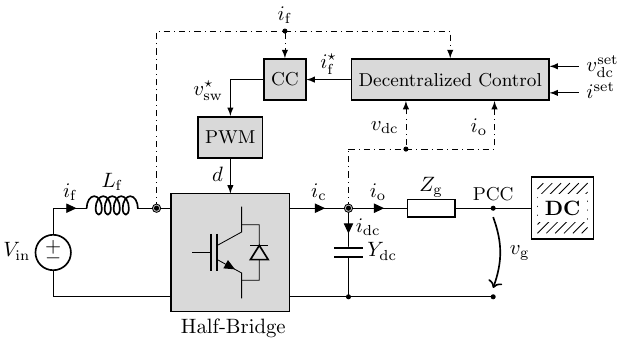}
    \vspace{-0.20cm}
    \caption{Illustration of a source DC/DC (bidirectional boost) converter connected to a DC microgrid and its associated control structure.}
    \label{fig:equivalent_circuit}
\end{figure}

An overview of the considered system is shown in Fig.~\ref{fig:equivalent_circuit}, where an exemplary DC/DC boost converter is connected to a DC microgrid. The converter is controlled by a conventional control architecture, illustrated using control blocks. The control scheme consists of a decentralized controller (e.g., droop or VDCM) that takes output voltage and current setpoints $v_\mathrm{dc}^\mathrm{set}$ and $i^\mathrm{set}$, respectively, and determines the filter current reference $i_\mathrm{f}^\star$. Subsequently, a PI current controller enforces the reference by establishing a voltage reference $v_\mathrm{sw}^\star$ at the converter output using PWM modulation. In the remainder of this section, we first describe the current controller and then outline several decentralized controller design approaches commonly employed in the literature.

\subsection{Current Control}
A central component of most DC/DC converter control schemes is the current controller, which regulates the filter current $i_\mathrm{f}$ to track the reference $i^\star_\mathrm{f}$ (output of the decentralized control layer) by generating the switching voltage reference:
\begin{equation}
    v_\mathrm{sw}^\star = R_\mathrm{i}(s)\bigl(i^\star_\mathrm{f} - i_\mathrm{f}\bigr).
\end{equation}
The PI current controller $R_\mathrm{i}(s)$ is tuned using an internal model control approach \cite{658735} and is given by:
\begin{equation}\label{eq:cc_bwdth}
    R_\mathrm{i}(s)
    =
    k_\mathrm{pi}\frac{1 + sT_\mathrm{ii}}{sT_\mathrm{ii}},\quad \,
    k_\mathrm{pi} = \omega_{\beta_\mathrm{i}} L_\mathrm{f},\quad \,
    T_\mathrm{ii} = \frac{C_\mathrm{dc} L_\mathrm{f}}{T_\mathrm{d}},
\end{equation}
where $L_\mathrm{f}$ denotes the filter inductance of the converter, $C_\mathrm{dc}$ is the output capacitance, and $T_\mathrm{d}$ represents the equivalent control delay accounting for sampling, computation, and PWM modulation effects. Additionally, $\omega_{\beta_\mathrm{i}}$ denotes the desired current-loop bandwidth and $T_\mathrm{ii}$ is the integral time constant. The bandwidth needs to be set sufficiently below the switching frequency $f_\mathrm{s}$, for example, at $7\%$ of the switching frequency, that is, $\omega_{\beta_\mathrm{i}} = 0.07 \cdot 2\pi f_\mathrm{s}$.

\subsection{Decentralized Control}
In this subsection, we review the most commonly employed decentralized control designs for power sharing and voltage control in DC microgrids.
\subsubsection{I--V Droop Control}
The considered I--V droop control law \cite{lazarevic_dc_microgrids} takes the following form:
\begin{equation}
    i_\mathrm{f}^\star = i^\mathrm{set} + 1/K_\mathrm{d}^\mathrm{dc} \left(G_\mathrm{lpf}(s)(v_\mathrm{dc}^\mathrm{set}-v_\mathrm{dc})\right), \label{eq:DC_droop}
\end{equation}
where $i_\mathrm{f}^\star$ denotes the calculated inductor current setpoint passed to the current controller, $i^\mathrm{set}$ represents the desired (nominal) current reference, $K_\mathrm{d}^\mathrm{dc}$ is the droop gain. The voltage difference between the measured voltage $v_\mathrm{dc}$ and its reference $v_\mathrm{dc}^\mathrm{set}$ determines the adjustment of the current reference, thus regulating the injection or absorption of power in a decentralized way. A low-pass filter $G_\mathrm{lpf}(s)$ is added to restrain the droop action to low-frequency voltage deviations and avoid amplification of high-frequency noise that may adversely affect system stability. 

\subsubsection{V--I Droop Control} 
In this droop control implementation, the droop gain is applied to a current deviation, creating a voltage reference $\Delta v_\mathrm{dc}^\star$ that is passed to a voltage controller:
\begin{equation}
    i_\mathrm{f}^\star = R_\mathrm{v}(s)\big(v^\mathrm{set}_\mathrm{dc}+\underbrace{Z_\mathrm{d}(s)(i^\mathrm{set}-i)}_{\Delta v_\mathrm{dc}^\star}-v_\mathrm{dc}\big), 
\end{equation}
where $R_\mathrm{v}(s)$ denotes the transfer function of the PI voltage controller. Several variants of V–I droop control exist. In the nominal formulation \cite{8096714}, hereinafter denoted by V--I\,$i_\mathrm{f}$, the feedback variable is the filter inductor current ($i = i_\mathrm{f}$), and a constant droop gain is employed, i.e., $Z_\mathrm{d}(s) = K_\mathrm{d}^\mathrm{dc}$. An alternative implementation \cite{CompareDroop2017} uses the output current ($i = i_\mathrm{o}$) as the feedback signal, referred to as V--I\,$i_\mathrm{o}$. Finally, \cite{liu2020} proposes replacing the constant droop gain with a frequency-dependent impedance $Z_\mathrm{d}(s)$ applied to the current deviation $i_\mathrm{o}^\mathrm{set} - i_\mathrm{o}$ to enhance transient performance. This configuration is denoted by V--I\,$Z_\mathrm{d}$.

\subsubsection{Virtual DC Machine}
The VDCM control leverages the mechanical characteristics of DC machines to emulate their rotational dynamics for voltage control, implemented as:
\begin{equation}
    i_\mathrm{f}^\star = 1/K_\mathrm{d}^\mathrm{dc}\left(R_\mathrm{dcm}(s)R_\mathrm{v}(s)(v^\mathrm{set}_\mathrm{dc}-v_\mathrm{dc})-v_\mathrm{dc}\right),
\end{equation}
where $R_\mathrm{dcm}(s)$ is a transfer function capturing the DC machine dynamics. The considered VDCM variant is the power-measurement-free VDCM scheme proposed in~\cite{Na2022PowerLoopFree}, which does not apply an explicit droop control in steady-state like the previous methods, but uses a PI voltage controller to track~$v^\mathrm{set}_\mathrm{dc}$.

\section{Source Converter Output Impedance Modeling} \label{sec:preliminaries}
The output impedance converter models are small‑signal representations in which the converter is modeled as a series connection of an internal voltage source and an equivalent impedance. Such models capture the input-output behavior of the converter, the dynamics of its passive components, and the influence of its control loops. We first introduce a generic small‑signal converter model and subsequently incorporate the control system to derive its output impedance.  

\subsection{Output Impedance Derivation}
Since converter dynamics are nonlinear, the system is first linearized around its operating point using state-space averaging, resulting in a small-signal equivalent model \cite{Erickson2001}. Assuming a constant input voltage, the inductor current $i_{\mathrm{f}}$ and converter terminal voltage $v_\mathrm{dc}$ deviations of a generic source converter are expressed as functions of the duty cycle $d$ and output current $i_\mathrm{o}$ perturbations, i.e.:
\begin{align}
    \Delta i_{\mathrm{f}} &= G_\mathrm{di}(s)\Delta d + G_\mathrm{oi}(s)\Delta i_\mathrm{o} , \label{eq:ilf_general} \\
    \Delta v_\mathrm{dc} &= G_\mathrm{dv}(s)\Delta d - Z_\mathrm{o}(s)\Delta i_\mathrm{o} , \label{eq:vdc_general}
\end{align}
where the operator $\Delta(\cdot)$ denotes small-signal perturbations around steady-state values. In these expressions, $G_\mathrm{di}(s)$ and $G_\mathrm{oi}(s)$ represent the transfer functions from $\Delta d$ and $\Delta i_\mathrm{o}$ to $\Delta i_{\mathrm{f}}$, respectively, while $G_\mathrm{dv}(s)$ denotes the transfer function from $\Delta d$ to $\Delta v_\mathrm{dc}$. In addition, $Z_\mathrm{o}(s)$ corresponds to the open-loop output impedance of the converter. For brevity, the explicit dependence on $(s)$ is omitted from this point onward.

For the specific case of the bidirectional boost converter analyzed in the remainder of this paper and depicted in Fig.~\ref{fig:equivalent_circuit}, the transfer functions in \eqref{eq:ilf_general} and \eqref{eq:vdc_general} can be written as:
\begin{subequations}
\begin{align}
    G_\mathrm{di} &= -\frac{I_\mathrm{o}}{D^2}\cdot\frac{1+\frac{Y_\mathrm{dc}V_\mathrm{o}}{I_\mathrm{o}}}{1+\frac{L_\mathrm{f}Y_\mathrm{dc}}{D^2}s},
    &G_\mathrm{oi} &= \frac{1}{D}\cdot\frac{1}{1+\frac{L_\mathrm{f}Y_\mathrm{dc}}{D^2}s},\\
    G_\mathrm{dv} &= -\frac{V_\mathrm{in}}{D^2}\cdot\frac{1-\frac{L_\mathrm{f}I_\mathrm{f}}{V_\mathrm{in}}s}{1+\frac{L_\mathrm{f}Y_\mathrm{dc}}{D^2}s},
    &Z_\mathrm{o} &= \frac{1}{D^2}\cdot\frac{L_\mathrm{f}s}{1+\frac{L_\mathrm{f}Y_\mathrm{dc}}{D^2}s},
\end{align}
\end{subequations}
where $Y_\mathrm{dc}=sC_\mathrm{dc}/(1+sC_\mathrm{dc}r_\mathrm{dc})$ denotes the equivalent admittance of the output capacitance, modeled as a capacitor in series with its internal resistance, and $L_\mathrm{f}$ represents the filter inductance. Furthermore, $I_\mathrm{o}$, $V_\mathrm{in}$, $V_\mathrm{o}$, and $D$ denote the steady-state output current, input voltage, output voltage, and duty cycle at the operating point, respectively. The duty cycle is defined as the relative on-time of the upper switch with respect to the switching period, while the inductor current is given by the well-known relationship $I_\mathrm{f}=I_\mathrm{o}/D$.

To derive the closed-loop output impedance, the converter's control system must be incorporated into the small-signal model. In particular, the duty-cycle perturbation $\Delta d$ is expressed in terms of the internal variables that serve as controller inputs, as follows:
\begin{equation}
    \Delta d = G_\mathrm{id}\Delta i_{\mathrm{f}} + G_\mathrm{od}\Delta i_\mathrm{o} + G_\mathrm{vd}\Delta v_\mathrm{dc} , \label{eq:duty_cycle}
\end{equation}
where $G_\mathrm{id}$, $G_\mathrm{od}$, and $G_\mathrm{vd}$ represent the respective transfer functions between the internal variables and $\Delta d$. These transfer functions depend on the selected decentralized control strategy and can be derived directly from the control relationships presented in the previous section.
In order to simplify the model, we eliminate $\Delta i_{\mathrm{f}}$ by substituting~\eqref{eq:ilf_general} into~\eqref{eq:duty_cycle}:
\begin{equation}
    \Delta d = \frac{G_\mathrm{id} G_\mathrm{oi} + G_\mathrm{od}}{1 - G_\mathrm{id} G_\mathrm{di}}  \Delta i_\mathrm{o} + \frac{G_\mathrm{vd}}{1 - G_\mathrm{id} G_\mathrm{di}}  \Delta v_\mathrm{dc} . \label{eq:duty_cycle_simplified}
\end{equation}
Finally, substituting~\eqref{eq:duty_cycle_simplified} into~\eqref{eq:vdc_general} and rearranging terms leads to the generalized expression of the source converter output impedance, i.e., the transfer function from $-\Delta i_\mathrm{o}$ to $\Delta v_\mathrm{dc}$:
\begin{equation}
\begin{split}
    Z_\mathrm{out} = \frac{Z_\mathrm{o}(1 - G_\mathrm{id} G_\mathrm{di}) - G_\mathrm{dv}(G_\mathrm{id} G_\mathrm{oi} + G_\mathrm{od})}
    {1 - G_\mathrm{id} G_\mathrm{di} - G_\mathrm{vd} G_\mathrm{dv}} . \label{eq:zout_generalized}
\end{split}
\end{equation}

\subsection{Alternative Small-Signal Representation} \label{sec:zout_prime}
The output voltage regulation of a source converter is closely linked to regulation of the current flowing into its output capacitance. During transient events, the current sharing between the converter and the output capacitor determines the resulting voltage deviation at the converter terminals~\cite{10621669}. This observation motivates an alternative small-signal representation of the source converter in which the capacitor dynamics are treated explicitly, while the remaining control-related dynamics are captured by an equivalent series output impedance.

\begin{figure}[t!]
    \centering
    \includegraphics[]{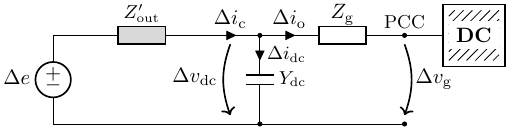}
    \vspace{-0.20cm}
    \caption{Small-signal circuit of a source converter connected to a DC microgrid.}
    \label{fig:equivalent_circuit_ss}
\end{figure}

With this in mind, the voltage control dynamics of a source converter connected to a DC microgrid are analyzed using the circuit shown in Fig.~\ref{fig:equivalent_circuit_ss}. The converter is modeled as a non-ideal voltage source with a series output impedance $Z_\mathrm{out}'$ and a parallel capacitive admittance $Y_\mathrm{dc}$, and is connected to the DC microgrid through the line impedance $Z_\mathrm{g}=r_\mathrm{g}+sL_\mathrm{g}$. Assuming a constant input voltage, i.e., $\Delta e=0$, this formulation is equivalent to the previously derived aggregated output impedance model, with $Z_\mathrm{out}=Z_\mathrm{out}'\parallel\frac{1}{Y_\mathrm{dc}}$. Thus, $Z_\mathrm{out}'$ denotes the series impedance, i.e., output impedance excluding the output capacitor admittance, and $Z_\mathrm{out}$ is the conventional output impedance (including the capacitor).

Based on Fig.~\ref{fig:equivalent_circuit_ss}, we can write the circuit equations describing the alternative model formulation as:
\begin{align}
    \Delta i_{\mathrm{dc}} &=Y_\mathrm{dc}\Delta v_\mathrm{dc} =  \Delta i_\mathrm{c} - \Delta i_\mathrm{o} , \label{eq:Cdc}\\
    \Delta v_\mathrm{dc}&= \Delta e - Z_\mathrm{out}'\Delta i_\mathrm{c} , \label{eq:vdc} \\
    \Delta i_{\mathrm{o}}&= \frac{\Delta v_\mathrm{dc} -\Delta v_\mathrm{g}}{Z_\mathrm{g}} \label{eq:idc} ,
\end{align}
where  $\Delta i_\mathrm{c}$ represents the converter current deviations at the converter output, $\Delta i_{\mathrm{dc}}$ denotes the current perturbations of the output capacitor, and $\Delta v_\mathrm{g}$ denotes the voltage perturbations at the point of common coupling (PCC) with the DC microgrid. Thus, in this model, the behavior of the voltage $\Delta v_\mathrm{dc}$ is directly influenced by $\Delta i_\mathrm{dc}$ and characteristics of impedance $Z_\mathrm{out}'$ as well as $C_\mathrm{dc}$, $Z_\mathrm{g}$, and $\Delta v_\mathrm{g}$. Under the assumption $\Delta e=0$, these relationships form the basis for characterizing the converter forming behavior, as discussed in the following section.

\section{Characterization of Forming Behavior} \label{sec:GFM_characterization}
In this section, we introduce three output impedance-based indices that quantify the converter’s ability to regulate its output voltage in the presence of output current and grid voltage variations. Each index characterizes a different aspect of the voltage control problem, and together they provide a comprehensive assessment of the voltage control performance.

\subsection{Definition of Performance Indices}
The output impedance itself serves as an indicator of the converter's voltage control performance. Intuitively, the smaller the magnitude of the impedance, the stiffer the output voltage, as the terminal voltage remains closer to the source voltage $\Delta e$ under disturbances. However, this intuition does not directly specify how \emph{small} the impedance should be to ensure satisfactory performance. To address this ambiguity, we introduce a normalized formulation of the output impedance.

\begin{definition}[Output Impedance Index]\label{def:OII}  The Output Impedance Index $\mathrm{OII}(\omega) \! : \! \Delta v_\mathrm{max}\mapsto \Delta v_\mathrm{dc}$ at angular frequency $\omega$ is defined as the transfer function mapping the disturbance in converter output current $-\Delta i_\mathrm{o}$ scaled by the voltage controller droop gain $K_\mathrm{d}$, i.e., $\Delta v_\mathrm{max}=-K_\mathrm{d}\Delta i_\mathrm{o}$ to the change in converter terminal voltage $\Delta v_\mathrm{dc}$:
\begin{equation}
\begin{aligned}
    \mathrm{OII}(\omega) &= \frac{Z_\mathrm{out}(j\omega)}{K_\mathrm{d}} \label{eq:OII} .
\end{aligned}
\end{equation}
\end{definition}
According to DC microgrid standards~\cite{dcindustrie2_2024}, the droop gain is selected as the ratio between the maximum permissible voltage deviation and the maximum allowed converter output current. Consequently, the droop gain has the same physical unit as the output impedance and establishes the upper bound on allowable disturbance amplification. Building on this relationship, the following classification provides the interpretation of the OII and clarifies how its values can be understood.

\begin{definition}[OII Converter Classification]\label{def:OII_classification}
Based on the magnitude of the $\mathrm{OII}$, the converter behavior can be classified~as:
\begin{equation*}
\mathrm{OII}(\omega)=
    \begin{cases}
        \|\mathrm{OII}(\omega)\|\leq 1, & \mathrm{grid\!-\!forming}, \\
        \|\mathrm{OII}(\omega)\|>1, & \mathrm{disturbance\!-\!amplifying}.
    \end{cases}
\end{equation*}
\end{definition}
This classification can be analyzed as follows. For $\|\mathrm{OII}(\omega)\|= 1 $, the voltage deviation at the converter terminal is proportional to the droop gain and output current deviations, i.e., in line with $\Delta v_\mathrm{max}$, while for $\|\mathrm{OII}(\omega)\|<1 $, the converter rejects the output current disturbances and forms a voltage by keeping absolute voltage deviations to less than $\Delta v_\mathrm{max}$. Thus, for $\|\mathrm{OII}(\omega)\| \leq1 $, the converter is said to possess grid-forming capabilities at the analyzed angular frequency $\omega$. In contrast, for $\|\mathrm{OII}(\omega)\| > 1 $, the converter amplifies the disturbances of the output current with a gain value higher than $K_\mathrm{d}$, thus operating in the disturbance-amplifying region. This leads to voltage overshoots and undershoots following changes in the output current~\cite{liu2020}, indicating undesired transient behavior.

Besides comparing the performance of control algorithms via the OII magnitude, the OII can also be used to assess stability. Specifically, passivity-based stability certificates can be reformulated in terms of the OII phase, rather than the phase of the output impedance~\cite{lazarevic_dc_microgrids}. One such certificate is presented in the following theorem. The proof is omitted, as it is well established for the output impedance \cite{BaoLee2007}.

\begin{theorem}[Passivity]
    The system in Fig.~\ref{fig:equivalent_circuit_ss} is stable if the following two conditions are satisfied:
    \begin{itemize}
        \item[\textit{(i)}] $\angle{\mathrm{OII}(\omega)}\in(-\pi/2,\pi/2),\forall \omega\in[0,\infty)$,
        \item[\textit{(ii)}] $\angle{Z_\mathrm{g}(\omega)}\in(-\pi/2,\pi/2),\forall \omega\in[0,\infty)$.
    \end{itemize}
\end{theorem}
That is, stable DC microgrid operation is ensured if the converter is passive, which requires the OII phase to remain within $(-90^\circ,\, 90^\circ)$, assuming that the grid itself consists solely of passive elements.

While the OII magnitude provides useful insights into the converter’s disturbance rejection capability, it does not fully describe how the converter interacts with the DC capacitance during transients. As highlighted in Section~\ref{sec:zout_prime}, terminal voltage regulation in DC microgrids is closely linked to the output capacitor current, since output current disturbances induce variations in the capacitor current, leading to voltage deviations at the converter terminals. Motivated by this observation, an additional index is introduced to characterize the current-forming behavior of the converter.

\begin{definition}[Current-Forming Index]\label{def:CFI} 
The Current-Forming Index $\mathrm{CFI}(\omega):\Delta i_\mathrm{o}\mapsto \Delta i_\mathrm{c}$ at angular frequency $\omega$ is defined as the transfer function mapping the disturbance in output current $\Delta i_\mathrm{o}$ to the change in converter current $\Delta i_\mathrm{c}$, which can be derived from \eqref{eq:Cdc} and $\Delta v_\mathrm{dc} = - Z_\mathrm{out} \Delta i_\mathrm{o}$ as:
\begin{equation}
\begin{aligned}
    \mathrm{CFI}(\omega) &= 1 - Y_\mathrm{dc}(j\omega)Z_\mathrm{out}(j\omega) \label{eq:CFI}.
\end{aligned}
\end{equation}
\end{definition}
The CFI is useful in characterizing the converter's current source behavior. Specifically, depending on the value of the CFI, the converter may act as a current source, rejecting the disturbances in the grid (output) current, follow the changes in the output current (as a voltage source), or amplify the disturbances in the output current. These behavior types are encompassed by the following classification.

\begin{definition}[CFI Converter Classification]\label{def:CFI_classification}
Based on the magnitude and phase of the $\mathrm{CFI}$, the converter behavior can be categorized by three different operating modes:
\begin{equation*}
    \mathrm{CFI}(\omega) =
    \begin{cases}
        \mathrm{CFI}(\omega)=1, & \mathrm{current\!-\!following},  \\
        \|\mathrm{CFI}(\omega)\|< 1, & \mathrm{current\!-\!forming},  \\
        \|\mathrm{CFI}(\omega)\|\geq1\footnotemark, & \mathrm{disturbance\!-\!amplifying}.  
    \end{cases}
\end{equation*}
\end{definition}
\footnotetext{Includes $\|\mathrm{CFI}(\omega)\| = 1$ only when $\angle \mathrm{CFI}(\omega) \neq 0$.}

For $\|\mathrm{CFI}(\omega)\| = 1$ and $\angle{\mathrm{CFI}(\omega)} = 0$, the converter current deviations $\Delta i_\mathrm{c}$ follow the output current deviations $\Delta i_\mathrm{o}$ at the frequency of interest, resulting in current-following operation. Consequently, $\Delta i_\mathrm{dc} = 0$, and the converter terminal voltage remains constant, as can be seen from~\eqref{eq:Cdc}. Similarly to the OII definition, for $\|\mathrm{CFI}(\omega)\|>1$ and $\|\mathrm{CFI}(\omega)\|= 1,\,\angle{\mathrm{CFI}(\omega)}\neq0$ the converter operates in the disturbance-amplifying region, since the internal converter current deviations $\Delta i_\mathrm{c}$ amplify the output current deviations $\Delta i_\mathrm{o}$, thus altering the capacitor current and terminal voltage of the converter.
Conversely, when $\|\mathrm{CFI}(\omega)\|<1$, the converter reacts less to the deviations of the output current and demonstrates current-forming capabilities. Consequently, the capacitor current $\Delta i_\mathrm{dc}$ follows the output current deviations with a higher magnitude of change, resulting in variations in DC voltage.

The two indices discussed thus far quantify the impact of disturbances in the output current $\Delta i_\mathrm{o}$. However, based on the equivalent circuit in Fig.~\ref{fig:equivalent_circuit_ss}, the PCC (i.e., grid) voltage perturbation $\Delta v_\mathrm{g}$ can alternatively be treated as the disturbance variable. This perspective reveals how variations in the grid voltage propagate to the converter terminal voltage and thereby motivates the definition of the third and last index.
\begin{definition}[Voltage-Forming Index]\label{def:VFI} The Voltage-Forming Index $\mathrm{VFI}(\omega):\Delta v_\mathrm{g}\mapsto \Delta v_\mathrm{dc}$ at angular frequency $\omega$ is defined as the transfer function mapping the disturbance in the PCC voltage $\Delta v_\mathrm{g}$ to the change in the converter voltage~$\Delta v_\mathrm{dc}$:
\begin{equation}
\begin{aligned}
    \mathrm{VFI}(\omega) &= \left(1+Z_{\mathrm{g}}(j\omega)Y_\mathrm{out}(j\omega)\right)^{-1} . \label{eq:VFI}
\end{aligned}
\end{equation}
\end{definition}
Analogous to the forming index introduced in \cite{zhuang2025} for AC systems, the VFI extends this concept to DC grids by characterizing the converter's voltage‑source behavior across different frequencies of interest. The following classification of VFI values aids the physical interpretation of the index.

\begin{definition}[VFI Converter Classification]\label{def:VFI_classification}
Based on the magnitude of the $\mathrm{VFI}$, the converter behavior can be classified~as:
\begin{equation*}
    \mathrm{VFI}(\omega)=
    \begin{cases}
        \|\mathrm{VFI}(\omega)\|\leq 1, & \mathrm{voltage\!-\!forming},  \\
        \|\mathrm{VFI}(\omega)\|>1, & \mathrm{disturbance\!-\!amplifying}.
    \end{cases}
\end{equation*}
\end{definition}

If $\|\mathrm{VFI}(\omega)\| \leq 1$, the converter imposes a voltage at its terminals either with the same amplitude as the grid voltage deviation for $\|\mathrm{VFI}(\omega)\| = 1$, or with a reduced amplitude for $\|\mathrm{VFI}(\omega)\| < 1$. This behavior is characteristic of voltage-forming operation at the angular frequency $\omega$. Conversely, if $\|\mathrm{VFI}(\omega)\| > 1$, the converter amplifies grid voltage deviations, thereby increasing the voltage fluctuations at its terminals and exhibiting disturbance-amplifying behavior.

\subsection{Desired Grid-Forming Converter Behavior}
Based on the previously defined indices, we now formalize the desired converter behavior, which we refer to as \emph{grid-forming behavior} to remain consistent with established terminology in AC networks. 

As discussed in Section~\ref{sec:zout_prime}, the output impedance $Z_\mathrm{out}$ of source converters can be represented as the parallel combination of the capacitive admittance $Y_\mathrm{dc}$ and series impedance $Z_\mathrm{out}'$. Following this formulation, $Z_\mathrm{out}'$ captures the converter’s control dynamics and determines its contribution to load variations through the droop gain $K_\mathrm{d}$, while $Y_\mathrm{dc}$ reduces terminal voltage deviations under such disturbances, i.e., when $\Delta i_\mathrm{c} \neq \Delta i_\mathrm{o}$, as described by \eqref{eq:Cdc}.
In this context, for a sufficiently large output capacitance, $Z_\mathrm{out}$ exhibits resistive (droop) behavior up to a certain frequency, beyond which capacitive behavior dominates and attenuates high-frequency disturbances, as detailed in \cite{liu2020}. The corresponding \emph{crossover frequency} $\omega_\mathrm{c}$ can be approximated by assuming $K_\mathrm{d} \gg r_\mathrm{dc}$:
\begin{equation}
    \omega_\mathrm{c}
    =
    \frac{1}{C_\mathrm{dc} K_\mathrm{d}}.
    \label{eq:fbd}
\end{equation}

However, in practice, $Z_\mathrm{out}'$ is shaped by control dynamics, delays, and converter non-idealities, while the output capacitance value is typically limited due to size and cost constraints. 
As a result, $Z_\mathrm{out}$ rarely exhibits purely resistive behavior up to $\omega_\mathrm{c}$, leading to $\|\mathrm{OII}(\omega)\| > 1$, i.e., disturbance-amplifying behavior according to the OII definition. In contrast, the desired OII behavior $\|\mathrm{OII}(\omega)\|\leq 1$ ensures attenuation of output current disturbances. Considering the finite control bandwidth of the converter and the size of its output capacitance, it is desirable to maintain $\|\mathrm{OII}(\omega)\| = 1$ for all $\omega \in [0,\,\omega_\mathrm{c}]$. This motivates an output impedance profile that is purely resistive up to $\omega_\mathrm{c}$ and becomes capacitive at higher frequencies, corresponding to voltage source behavior. Following this reasoning, we formally define the \emph{desired converter output impedance} as
\begin{equation}
    Z_\mathrm{out,d}(j\omega)
    =
    \frac{K_\mathrm{d}}{1 + Y_\mathrm{dc}(j\omega)K_\mathrm{d}},
    \label{eq:Z_desired}
\end{equation}
which represents a parallel connection of a resistive branch $Z_\mathrm{out,d}'(j\omega)=K_\mathrm{d}$, governing low- and mid-frequency behavior, and the output capacitance, which dominates for $\omega > \omega_\mathrm{c}$, where $\omega_\mathrm{c}$ is given by \eqref{eq:fbd}.

The relationship between the desired behavior and the CFI is revealed through the analytical connection between the OII and the CFI, obtained from \eqref{eq:OII} and \eqref{eq:CFI}:
\begin{equation}
    \mathrm{OII}(\omega)
    =
    \frac{Z_\mathrm{out}'(j\omega)}{K_\mathrm{d}} \,\mathrm{CFI}(\omega).
    \label{eq:OII_CFI}
\end{equation}
The relationship illustrates the equivalence between the operating regions characterized by the OII and CFI in the idealized case where $Z_\mathrm{out}'(j\omega)=K_\mathrm{d}$. It follows that both the low-frequency current-following and high-frequency current-forming behavior defined by the CFI correspond to the desired grid-forming operation, as defined in Definitions~\ref{def:CFI_classification}~and~\ref{def:OII_classification}, respectively. In the low-frequency current-following region, the converter forms the voltage based on its output impedance excluding the output capacitance, i.e., $Z_{\mathrm{out}}'$. In contrast, in the high-frequency current-forming region, the voltage is formed primarily through the output capacitance of the converter, and therefore corresponds to the equivalent output impedance $Z_{\mathrm{out}}$.

Interpreting the VFI is more challenging, as it is defined at the system level. Specifically, unlike the OII and CFI, which depend solely on converter (device-level) parameters, the VFI also depends on the grid impedance $Z_\mathrm{g}$. As a result, the VFI is influenced not only by the converter itself, but also by its interaction with the grid, making it a system-level index. This relationship becomes clearer in the following results section.

\section{Results} \label{sec:results_freq}

\begin{figure}[!b]
	\centering
    \vspace{-0.30cm}
	\includegraphics[width=0.485\textwidth]{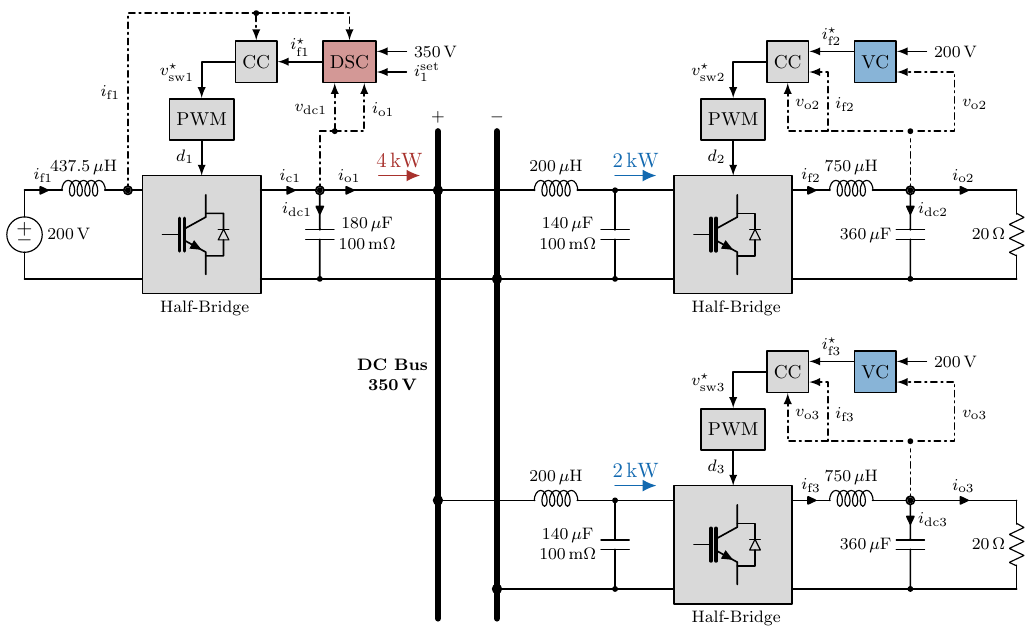}
	\vspace{-0.60cm}
    \caption{DC microgrid with a boost-type source converter (left) supplying two buck-type constant power load converters (right). The source converter is controlled using decentralized control (DSC) algorithms described in Section~\ref{sec:control_algorithms}, while the load converter is controlled using a PI output voltage controller (VC).}
	\label{fig:microgrid_example}
\end{figure}

\subsection{System Setup}
In order to analyze the grid-forming behavior of legacy decentralized control architectures from the literature, we analyze a simple microgrid setup as illustrated in Fig.~\ref{fig:microgrid_example}. The considered microgrid consists of a source boost converter and two load buck converters. The boost converter is controlled with the different control algorithms analyzed in Section~\ref{sec:control_algorithms} and is responsible for stabilizing the DC bus voltage. The two buck converters are controlled to keep the voltage on the load side constant, exhibiting constant power load behavior. 

\subsection{Frequency-Domain Analysis}
\begin{figure*}[t!]
    \centering
    \includegraphics[width=\textwidth]{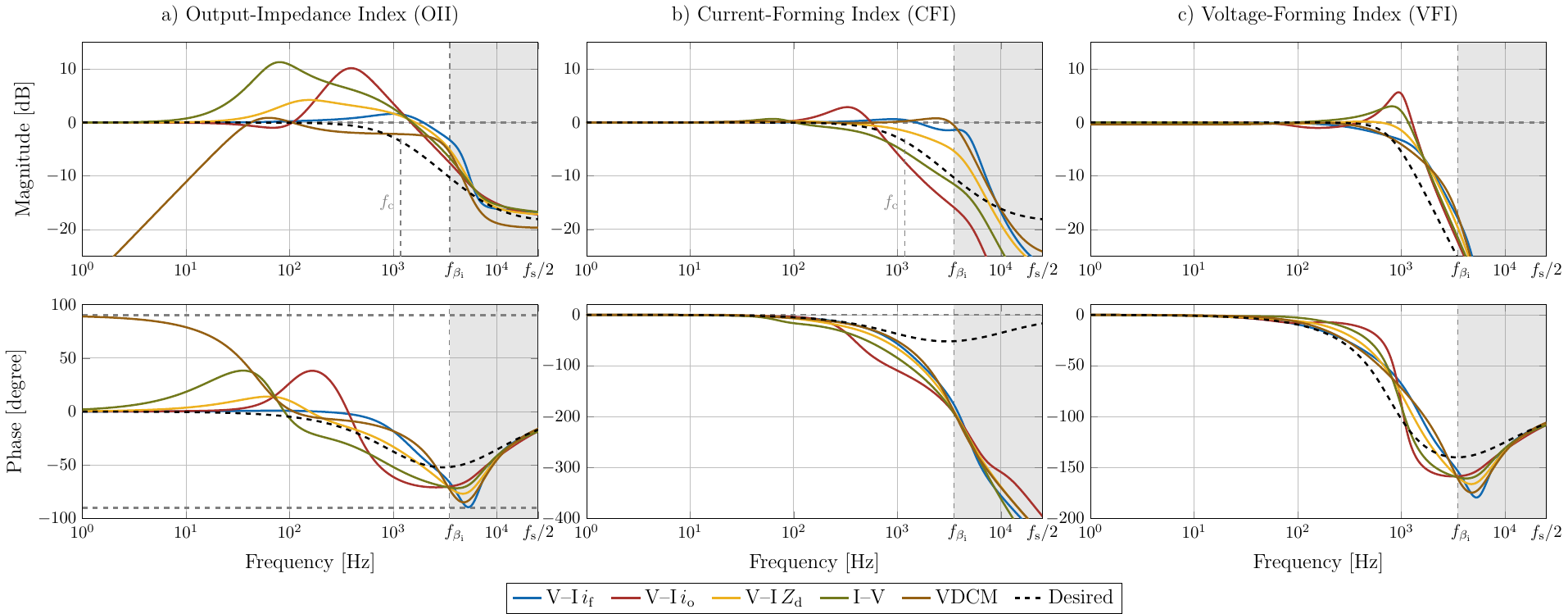}
    \vspace{-0.40cm}
    \caption{Converter forming characterization indices for different boost converter control algorithms: a) Output-Impedance Index (OII), b) Current-Forming Index (CFI), and c) Voltage-Forming Index (VFI). The grey shaded area corresponds to the frequency range above the current-control bandwidth $f_{\beta_\mathrm{i}}$.}
    \label{fig:index_plots}
\end{figure*}

First, we analyze the device-level behavior of the source converter under the different control algorithms using the forming indices introduced in Section~\ref{sec:GFM_characterization} and shown in Fig.~\ref{fig:index_plots}. The trajectories indicate that some algorithms approximate the desired behavior over limited frequency ranges, but none achieve it across the full spectrum. Specifically, the V--I\,$i_{\mathrm{f}}$ and VDCM controllers exhibit predominantly grid-forming behavior, maintaining $\|\mathrm{OII}(\omega)\| \leq 1$ over most frequencies and largely avoiding the disturbance-amplifying region, as seen in Fig.~\ref{fig:index_plots}a. In contrast, the remaining algorithms operate in the disturbance-amplifying region over a wider frequency range. These trends are consistent with the CFI and VFI plots from Figs.~\ref{fig:index_plots}b--c, where the best-performing controllers operate mainly in the current-following and forming regions at low and high frequencies, while exhibiting voltage-forming behavior.

\begin{figure}[t!]
	\centering
	\includegraphics[width=0.475\textwidth]{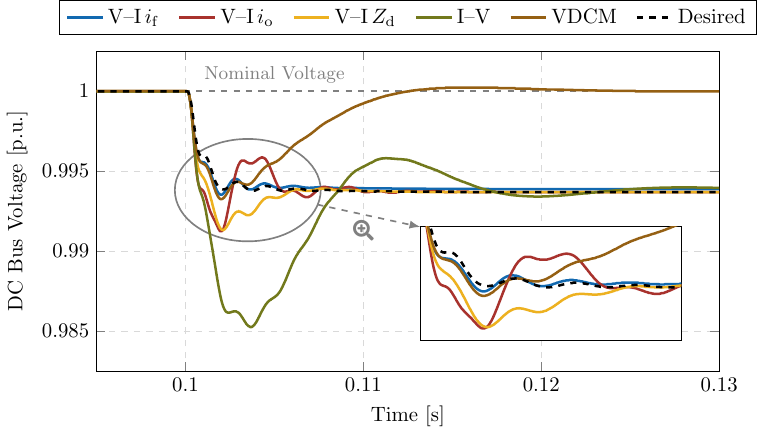}
	\vspace{-0.20cm}
    \caption{DC bus voltage response to a load increase for different boost converter control algorithms under analysis.}
	\label{fig:results_simulations}
\end{figure}

\subsection{Time-Domain Analysis}
To verify the analytical findings from the forming indices, we simulate the source converter response to a disturbance in \textsc{Matlab}/\textsc{Simulink}. At $0.1\,$s simulation time, the load of one buck converter is doubled, and the resulting DC bus voltage is observed for the different control algorithms. For reference, the boost converter is also replaced with a voltage source modeled using the desired output impedance from \eqref{eq:Z_desired}.

The DC bus voltage trajectories are shown in Fig.~\ref{fig:results_simulations}, linking the analytical findings to the dynamic response of each algorithm: the V--I\,$i_\mathrm{f}$ and VDCM controllers, which most closely match the desired impedance behavior, exhibit the best disturbance response and approximate the desired voltage source behavior. Moreover, the remaining algorithms show responses consistent with their OII trajectories in Fig.~\ref{fig:index_plots}a, highlighting a clear correlation: larger $\|\mathrm{OII}(\omega)\|$ overshoot (operation above unity gain) results in a more pronounced voltage undershoot.

\section{Conclusion} \label{sec:concl}
This paper introduces three novel output-impedance–based indices for quantifying the forming behavior of source converters in DC microgrids. The proposed indices evaluate converter behavior with respect to both output voltage and output current, requiring no detailed knowledge of the microgrid.  Thus, they provide a unified and consistent framework for comparing different converter-control algorithms and can further be used to tune existing droop controllers or design new control schemes.

\bibliographystyle{IEEEtran}
\bibliography{bibliography}

\end{document}